\documentclass[aps,prb,twocolumn,letterpaper,superscriptaddress,showpacs]{revtex4-1}
\usepackage{CJK}
\usepackage{keyval}%
\usepackage{graphicx}
\usepackage{dcolumn}
\usepackage{bm}
\usepackage{color}
\usepackage{amsmath, amssymb}
\usepackage{soul}

\setkeys{Gin}{width=0.8\columnwidth,clip=true}
\newcommand{\ignore}[1]{}

\begin{document}

\begin{CJK*}{UTF8}{bsmi}
\title{Fermi surface interconnectivity and topology in Weyl fermion semimetals TaAs, TaP, NbAs, and NbP}
\author{Chi-Cheng Lee}
\affiliation{Centre for Advanced 2D Materials and Graphene Research Centre National University of Singapore,
6 Science Drive 2, Singapore 117546}%
\affiliation{Department of Physics, National University of Singapore, 2 Science Drive 3, Singapore 117542}%

\author{Su-Yang Xu}\affiliation {Laboratory for Topological Quantum Matter and Spectroscopy (B7), Department of Physics, Princeton University, Princeton, New Jersey 08544, USA}

\author{Shin-Ming Huang}
\affiliation{Centre for Advanced 2D Materials and Graphene Research Centre National University of Singapore,
6 Science Drive 2, Singapore 117546}%
\affiliation{Department of Physics, National University of Singapore, 2 Science Drive 3, Singapore 117542}%

\author{Daniel S. Sanchez\footnote{These authors contributed equally to this work.}}\affiliation {Laboratory for Topological Quantum Matter and Spectroscopy (B7), Department of Physics, Princeton University, Princeton, New Jersey 08544, USA}
\author{Ilya Belopolski}\affiliation {Laboratory for Topological Quantum Matter and Spectroscopy (B7), Department of Physics, Princeton University, Princeton, New Jersey 08544, USA}

\author{Guoqing Chang}
\affiliation{Centre for Advanced 2D Materials and Graphene Research Centre National University of Singapore,
6 Science Drive 2, Singapore 117546}%
\affiliation{Department of Physics, National University of Singapore, 2 Science Drive 3, Singapore 117542}%

\author{Guang Bian}\affiliation {Laboratory for Topological Quantum Matter and Spectroscopy (B7), Department of Physics, Princeton University, Princeton, New Jersey 08544, USA}
\author{Nasser Alidoust}\affiliation {Laboratory for Topological Quantum Matter and Spectroscopy (B7), Department of Physics, Princeton University, Princeton, New Jersey 08544, USA}
\author{Hao Zheng}\affiliation {Laboratory for Topological Quantum Matter and Spectroscopy (B7), Department of Physics, Princeton University, Princeton, New Jersey 08544, USA}
\author{Madhab Neupane}\affiliation {Laboratory for Topological Quantum Matter and Spectroscopy (B7), Department of Physics, Princeton University, Princeton, New Jersey 08544, USA}
\affiliation {Condensed Matter and Magnet Science Group, Los Alamos National Laboratory, Los Alamos, NM 87545, USA}

\author{Baokai Wang}
\affiliation{Centre for Advanced 2D Materials and Graphene Research Centre National University of Singapore,
6 Science Drive 2, Singapore 117546}%
\affiliation{Department of Physics, National University of Singapore, 2 Science Drive 3, Singapore 117542}%
\affiliation{Department of Physics, Northeastern University, Boston, Massachusetts 02115, USA}

\author{Arun Bansil}
\affiliation{Department of Physics, Northeastern University, Boston, Massachusetts 02115, USA}

\author{M. Zahid Hasan\footnote{Corresponding authors (emails): mzhasan@princeton.edu}}\affiliation {Laboratory for Topological Quantum Matter and Spectroscopy (B7), Department of Physics, Princeton University, Princeton, New Jersey 08544, USA}
\affiliation{Princeton Center for Complex Materials, Princeton Institute for the Science and Technology of Materials, Princeton University, Princeton, New Jersey 08544, USA}

\author{Hsin Lin}
\affiliation{Centre for Advanced 2D Materials and Graphene Research Centre National University of Singapore,
6 Science Drive 2, Singapore 117546}%
\affiliation{Department of Physics, National University of Singapore, 2 Science Drive 3, Singapore 117542}%

\date{\today}

\begin{abstract}
The family of binary compounds including TaAs, TaP, NbAs, and NbP was recently discovered as the first realization of Weyl semimetals. In order to develop a comprehensive description of the charge carriers in these Weyl semimetals, we performed a detailed and systematic electronic band structure calculations which reveal the nature of Fermi surfaces and their complex interconnectivity in TaAs, TaP, NbAs, and NbP.  Our work report the first comparative and comprehensive study of Fermi surface topology and band structure details of all known members of the Weyl semimetal family and provide the necessary building blocks for advancing our understanding of their unique topologically protected low-energy Weyl fermion physics.
\end{abstract}

\pacs{}

\maketitle
\end{CJK*}

\section{Introduction}

The discovery of Dirac fermions as low-energy quasi-particle excitations in graphene and on the surfaces of topological insulators has drawn significant attention in both fundamental physics research and device applications \cite{Weyl, herring_accidental_1937, abrikosov_properties_1971, volovik_universe_2009,Murakami, PhysRevB.83.205101,PhysRevB.84.075129,Balents,PhysRevLett.107.127205, RevModPhys.81.109, RevModPhys.83.407, RevModPhys.82.3045, RevModPhys.83.1057, 2014arXiv1401.0529H, PhysRevB.86.115208, 2013arXiv1301.0330T, PhysRevLett.107.186806,PhysRevB.89.081106, PhysRevB.90.155316, PhysRevLett.108.046602, PhysRevLett.111.246603, PhysRevX.4.031035, hosur2013recent, nielsen1983adler, potter2014quantum, fang2003anomalous, PhysRevB.85.035103, PhysRevB.85.165110, PhysRevB.88.035444, PhysRevB.87.245112,huang_weyl_2015, PhysRevX.5.011029, 2015arXiv150203807X, 2015arXiv150204684L}. Since these Dirac-like fermions propagate as massless relativistic particles they behave differently from the conventional charge carriers in metals, semiconductors, and insulators. Recently, a new form of massless fermions with lifted degeneracy at the nodal point, from four-fold to two-fold, has been proposed to exist in condensed matter systems through a time reserval or inversion symmetry breaking mechanism. This symmetry breaking operation modifies the quasi-particle's dispersion relation from a Dirac to a Weyl equation.\cite{Weyl, herring_accidental_1937, abrikosov_properties_1971, volovik_universe_2009,Murakami, PhysRevB.83.205101,PhysRevB.84.075129,Balents,PhysRevLett.107.127205, RevModPhys.81.109, RevModPhys.83.407, RevModPhys.82.3045, RevModPhys.83.1057, 2014arXiv1401.0529H, PhysRevB.86.115208, 2013arXiv1301.0330T, PhysRevLett.107.186806,PhysRevB.89.081106, PhysRevB.90.155316, PhysRevLett.108.046602, PhysRevLett.111.246603, PhysRevX.4.031035, hosur2013recent, nielsen1983adler, potter2014quantum, fang2003anomalous, PhysRevB.85.035103, PhysRevB.85.165110, PhysRevB.88.035444, PhysRevB.87.245112,huang_weyl_2015, PhysRevX.5.011029, 2015arXiv150203807X, 2015arXiv150204684L} Many theoretical proposals exist for realizing Weyl semimetals that possess interesting physical properties, such as discontinuous Fermi arcs and negative magnetoresistance due to the chiral anomaly.\cite{Balents,PhysRevB.83.205101,PhysRevB.84.075129,2014arXiv1401.0529H,PhysRevB.86.115208,Murakami,2013arXiv1301.0330T,PhysRevLett.107.127205,PhysRevLett.107.186806,PhysRevB.89.081106,PhysRevB.90.155316,PhysRevB.85.035103,PhysRevB.85.165110,PhysRevB.87.245112,nielsen1983adler,PhysRevLett.108.046602,PhysRevLett.111.246603,hosur2013recent,PhysRevX.4.031035,potter2014quantum} A Weyl node with definite chirality is associated with the Berry curvature and may be thought of as realizations of magnetic monopoles in momentum space.\cite{berry1984quantal,fang2003anomalous}

Recent theoretical works have proposed the realization of the Weyl semimetal state in the inversion symmetry breaking TaAs family.\cite{huang_weyl_2015,PhysRevX.5.011029} Shortly after the prediction, the first Weyl semimetal was experimentally discovered in TaAs \cite{2015arXiv150203807X}. The electronic Weyl semimetal state in TaAs was experimentally observed through photoemission spectroscopy\cite{2015arXiv150203807X,2015arXiv150204684L}. Electrical transport experiments have shown that TaAs has very high mobility \cite{2015arXiv150200251Z} and, therefore, is consistent with the protected nature of the Weyl fermions and reported signatures of the chiral anomaly \cite{2015arXiv150301304H,2015arXiv150302630Z}. Soon after the initial experimental discovery other independent photoemission experimental works have confirmed the  Weyl  semimetal state  in  TaAs  and, furthermore,  the  Weyl  state  in other members of the same family, which  includes NbAs and TaP. \cite{2015arXiv150309188L,2015arXiv150401350X,2015arXiv150700521Y,2015arXiv150703983X,2015arXiv150803102X}

%The binary Weyl semimetals, TaAs, TaP, NbAs, and NbP, have been theoretically
%and experimentally investigated very recently.\cite{huang_weyl_2015,PhysRevX.5.011029,2015arXiv150200251Z,2015arXiv150203807X,2015arXiv150204684L,2015arXiv150302630Z,2015arXiv150301304H,ghimire2015magnetotransport,2015arXiv150309188L,2015arXiv150204361S,2015arXiv150401350X,2015arXiv150601751L,2015arXiv150606577S}
Provided the inclusion of spin-orbit coupling, these four non-centrosymmetric compounds are proposed to realize Weyl fermions without breaking time-reversal symmetry and, consequently, are easily studied under ambient conditions. For example, no external magnetic field or pressure is required for the realization of the Weyl semimetal state in this TaAs family.\cite{huang_weyl_2015,2015arXiv150203807X}
Conventional semimetals are materials that have a small overlap in energy between the valence and conduction bands. By contrast, a Weyl semimetal has valence and conduction bands that touch at a set of discrete points in the bulk Brillouin zone, called the Weyl nodes. In the presence of additional doping, Fermi pockets, or multiple Weyl nodes at different energies, a Weyl semimetal will naturally contain electron and hole charge carriers, giving rise to a compensated semimetal. This is indeed the case for the TaAs family of Weyl materials as reported in photoemission and transport experiments \cite{2015arXiv150204684L, 2015arXiv150703983X, 2015arXiv150601751L, shekhar_extremely_2015, 2015arXiv150203807X,2015arXiv150401350X,2015arXiv150803102X,2015arXiv150200251Z,2015arXiv150301304H,2015arXiv150302630Z,ghimire2015magnetotransport,2015arXiv150600924W,2015arXiv150606577S,2015arXiv150706981M,2015arXiv150706301Z,2015PhRvB..92d1203Z,2015arXiv150705246D, 2015arXiv150309188L, 2015arXiv150700521Y}
%(\textbf{please cite our ARPES papers on TaAs, NbAs,and TaP and also these transport http://arxiv.org/abs/1502.00251, http://arxiv.org/abs/1502.04361 (10.1038/nphys3372), http://arxiv.org/abs/1503.01304, http://arxiv.org/abs/1503.02630, http://arxiv.org/abs/1503.07571 (J. Phys.: Condens. Matter 27 (2015) 152201), http://arxiv.org/abs/1506.00924, http://arxiv.org/abs/1506.06577, http://arxiv.org/abs/1507.06981, http://arxiv.org/abs/1507.06301, http://arxiv.org/abs/1507.01298}, http://arxiv.org/abs/1507.05246). 
The Fermi surfaces can originate from the conduction or valence band crossing the Fermi energy, the Weyl cones, and their combinations. The charge carriers are expected to have small concentrations with a high mobility, at least for the Weyl fermions. The interplay between tunable concentrations of multiple charge carriers, high Fermi velocities, and the protected nature of the Weyl fermion carriers suggests a new avenue for designing novel electronic devices. The Weyl semimetals are also good candidates for exploring critical phenomena near the Fermi energy, due to their small carrier concentrations and linear band touchings. Therefore, it is of interest to explore the electronic structure of these four compounds to better understand their similarities and differences. 

Previous theoretical studies of this family focused on establishing TaAs as a Weyl semimetal and therefore only reported a few key aspects of the electronic structure for that purpose. On the other hand, many systematic details of the band structure, which are crucial for spectroscopy and transport studies, have not been studied. Furthermore, a comparative study of the electronic structure of TaAs with the other members remains lacking. In this study, we perform a comprehensive first-principles calculation to explore the properties of Fermi surfaces for the entire family of Weyl semimetals, TaAs, TaP, NbAs, and NbP, and, thereby, providing the necessary building block for future investigations of Weyl semimetals from a theoretical, experimental and application perspective. The paper is organized in the following manner:  In Section~\ref{sec:computational} we provide the computational details. In Section~\ref{sec:bandstructure}, electronic band structures and density of states are studied to improve our basic understanding of these four compounds. In Section~\ref{sec:fermisurface}, we discuss the details of the Fermi velocities of Weyl fermions and Fermi surfaces. Finally, in Section~\ref{sec:discussion} we present a brief discussion covering the important conclusions reached.  

\section{Computational detail}

\label{sec:computational}

\begin{figure}[tbp]
\includegraphics[width=1.00\columnwidth,clip=true,angle=0]{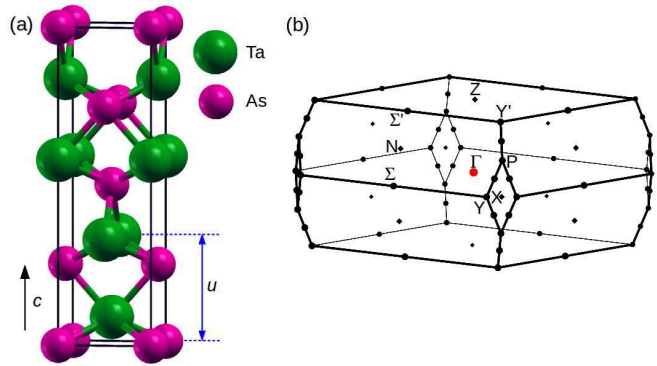}
\caption{(Color online). (a) Conventional unit cell of TaAs. (b) Primitive first Brillouin
zone of TaAs. TaP, NbAs, and NbP share the same body-centered-tetragonal
structure with different lattice parameters listed in Table~\protect\ref%
{table:lattice}. }
\label{fig:structure}
\end{figure}

First-principles calculations of TaAs, TaP, NbAs, and NbP, which have space group $I4_{1}md$ (109), a body-centered-tetragonal structure, were performed using the OpenMX code. The OpenMX code was based on norm-conserving pseudopotentials generated with multi reference energies\cite{MBK} and optimized pseudoatomic basis functions.\cite{Ozaki,openmx} For each Ta atom, three, two, two, and one optimized radial functions were allocated for the $s$-, $p$-, $d$-, and $f$-orbitals ($s3p2d2f1$), respectively. For As, Nb, and P atoms, $s3p3d3f2$, $s3p3d3f1$, and $s3p3d2f1$ were adopted, respectively. A cut-off radius of 7 Bohr was adopted for Ta, Nb, and P basis functions while a 9 Bohr cut-off radius was used for the As basis functions. The spin-orbit coupling was taken into account through $j$-dependent pseudopotentials,\cite{PhysRevB.64.073106} and the generalized gradient approximation (GGA) was adopted for the exchange-correlation energy functional.\cite{Kohn,Perdew} The cut-off energy of 1000 Ry was used for numerical integrations and for the solution of the Poisson equation. For the quality of  $k$-point sampling, a $17\times 17\times 5$ mesh for the conventional unit cell was adopted. The experimental lattice parameters were chosen in the calculations and are listed in Table~\ref{table:lattice}. The conventional unit cell and first Brillouin zone of the primitive unit cell of the representative TaAs is plotted in Fig.~\ref{fig:structure}.

Throughout this paper, for brevity, we will label the crystal momentum in units of $(2\pi/a,2\pi/a,2\pi/c)$ unless otherwise specified.

\begin{table}[tbp]
\caption{Experimental lattice parameters ($cf.$ Fig.~\protect\ref%
{fig:structure}).}
\label{table:lattice}%
\begin{tabular}{ccccc}
\hline\hline
& TaAs & TaP & NbAs & NbP \\ \hline
$a \ \text{(\AA)}$ & 3.437   & 3.318  & 3.452  & 3.334
 \\
$c \ \text{(\AA)}$ & 11.656  & 11.363  & 11.679  & 11.376  \\ \hline
$u/c$
 & 0.333 & 0.334 & 0.333 & 0.334 \\ \hline\hline
\end{tabular}%
\end{table}

\section{Electronic structure}

\label{sec:bandstructure}

In this section, the electronic band structure and density of states for the four Weyl semimetal compounds are presented. With the combination of Ta, or Nb, and As, or P, very similar electronic structures can be found and, therefore, illustrates a flexibility in realizing Weyl fermions among these four elements.

\subsection{Symmetry}

The relevant symmetry elements in the TaAs family (space group $I4_{1}md$) is the fourfold screw rotations along $z$-axis and two mirror reflections with respect to $x$- and $y$-axis. Due to lack of inversion symmetry, the rotation axis cannot sit perpendicular to the mirror planes. The combination of screw rotations and mirror reflections results in diagonal glide planes. Because the screw rotation and glide reflection are compound symmetry operations of translation and point group operations, for specific surface termination the surface states will not contain these compound symmetries. For the bulk, however, due to translation symmetry, the energy spectrum displays the $C_{4}$ and mirror symmetries. In $k$ space, there are mirror planes at $k_{x}$%
=0, $k_{y}$=0 and $k_{x}\pm k_{y}$=0. The $k_{x}$=0 and $k_{y}$=0 mirror planes are important for the realization of Weyl nodes. On the mirror plane, each band can possess a mirror eigenvalue of either  $+1$ or $-1$ ($+i$ or $-i$) in the absence (presence) of spin-orbit coupling, respectively. Bands with opposite mirror eigenvalues are allowed to cross, and those with the same mirror eigenvalues are only permitted to anti-cross. The other symmetry is time-reversal, which is not a spatial symmetry. Since time-reversal symmetry relates $\vec{k}$ and $-\vec{k}$ states, its combination with the twofold rotation results in the energy spectrum being symmetric with respect to the $k_{z}=0, 1$ planes.

Since a Weyl node is characterized by a chirality of $\pm1$, which is defined by the integration of Berry curvature, a pseudovector, over a closed $k$-space surface (in unit of $2\pi$),\cite{berry1984quantal,fang2003anomalous}, Weyl nodes related by mirror symmetry will possess an opposite chirality while those related by rotation and time-reversal symmetries will have the same chirality. Thus, a mirror pair of Weyl nodes, as mirror images, will possess, relative to each other, an opposite chirality, and two Weyl nodes at equal $k_x$ and $k_y$, but opposite $k_z$, will have the same chirality. Furthermore, the closed surface can be extended to a Fermi surface, yielding a Chern number for the Fermi surface\cite{2015arXiv150507727G} and, therefore, reflecting the net chirality of the enclosed Weyl nodes.

\subsection{Band structure}

\begin{figure}[tbp]
\includegraphics[width=0.87\columnwidth,clip=true,angle=0]{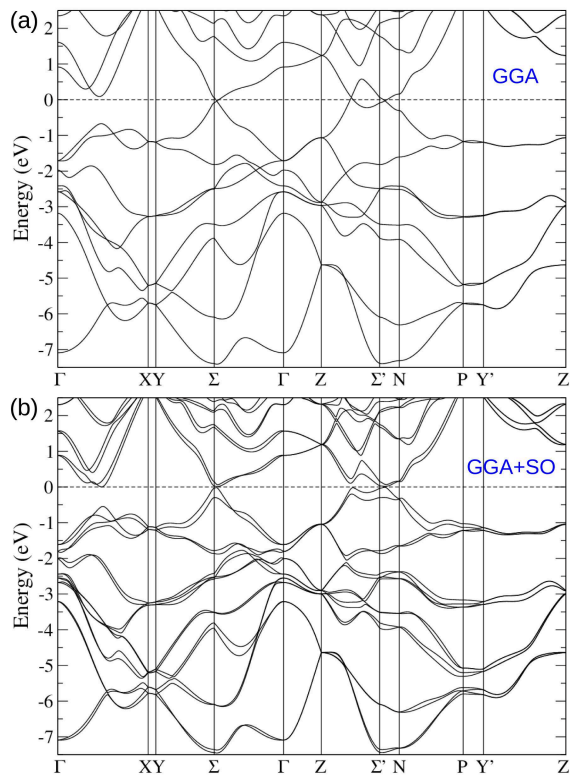}
\caption{Electronic band structure of TaAs (a) without and (b) with
spin-orbit coupling. }
\label{fig:bandtaas}
\end{figure}

\begin{figure}[tbp]
\includegraphics[width=0.87\columnwidth,clip=true,angle=0]{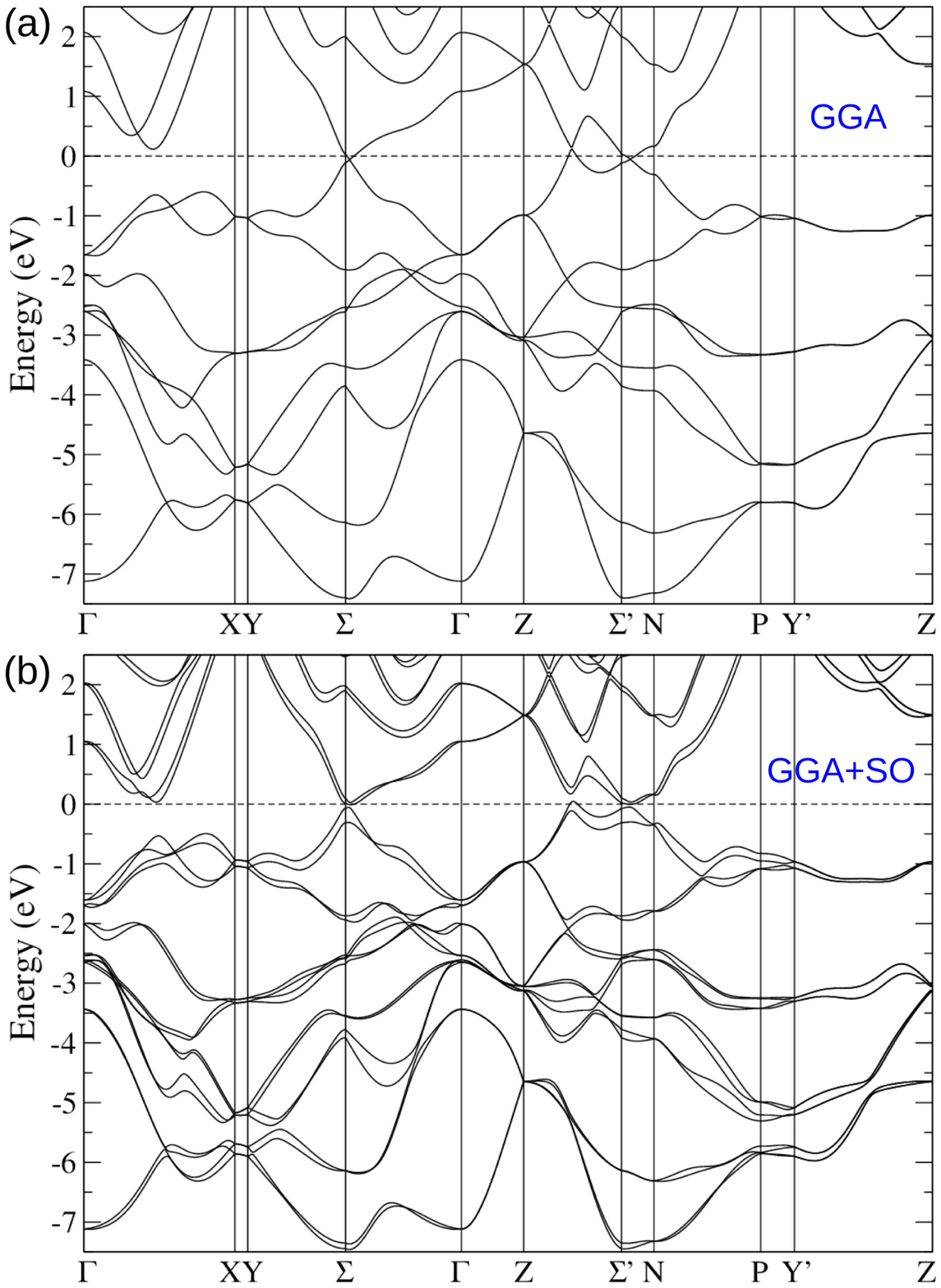}
\caption{Electronic band structure of TaP (a) without and (b) with
spin-orbit coupling. }
\label{fig:bandtap}
\end{figure}

\begin{figure}[tbp]
\includegraphics[width=0.87\columnwidth,clip=true,angle=0]{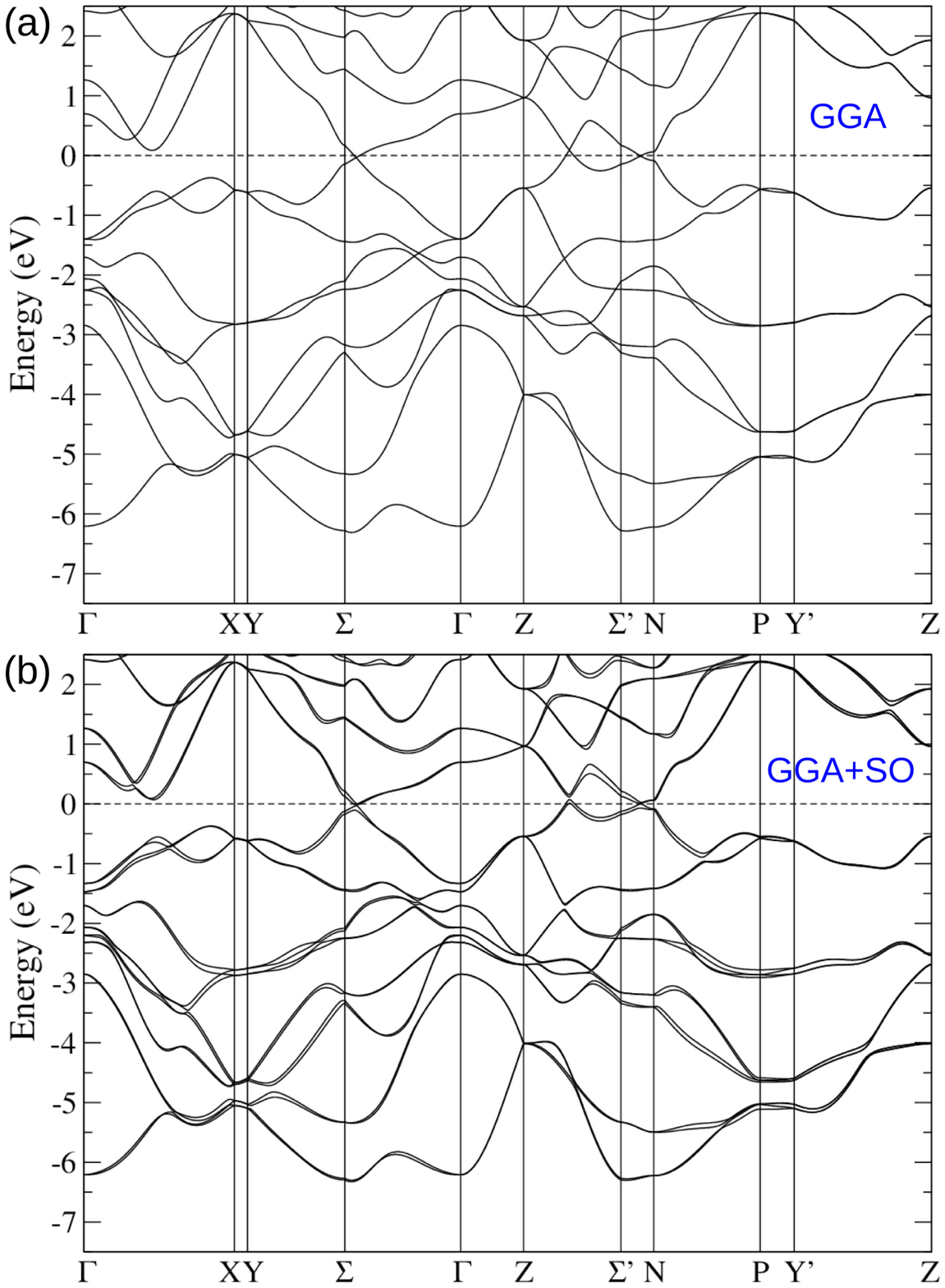}
\caption{Electronic band structure of NbAs (a) without and (b) with
spin-orbit coupling. }
\label{fig:bandnbas}
\end{figure}

\begin{figure}[tbp]
\includegraphics[width=0.87\columnwidth,clip=true,angle=0]{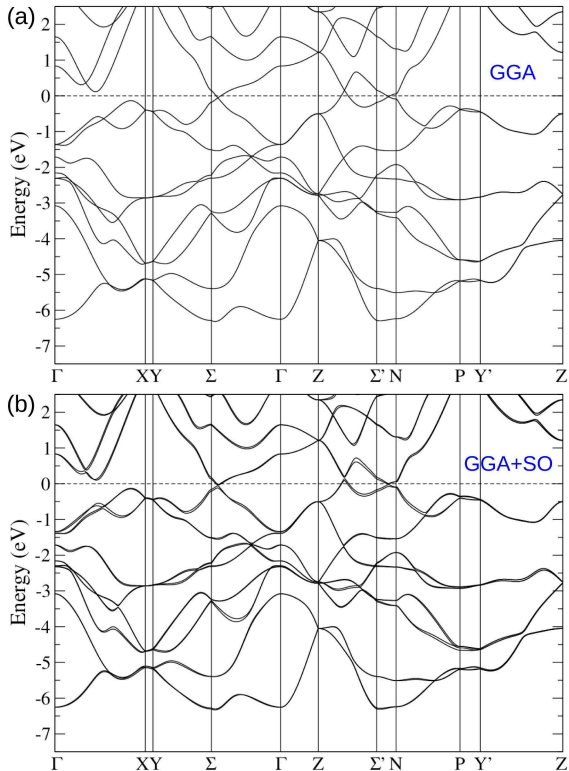}
\caption{Electronic band structure of NbP (a) without and (b) with
spin-orbit coupling. }
\label{fig:bandnbp}
\end{figure}

\begin{figure}[tbp]
\includegraphics[width=1.00\columnwidth,clip=true,angle=0]{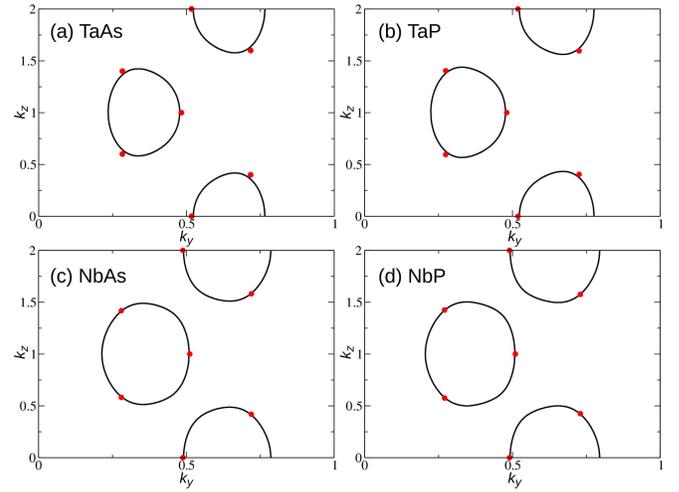}
\caption{(Color online). Line nodes on the $k_x$=0 plane formed by the crossing of valence and conduction bands
in (a) TaAs, (b) TaP, (c) NbAs, and (d) NbP without spin-orbit coupling. The
red solid circles indicated the projection of Weyl nodes on the $k_x$=0 plane after turning on the spin-orbit coupling. For each circle, two Weyl nodes of opposite chiralities on two sides of the $k_x$=0 plane are found.}
\label{fig:loops}
\end{figure}

The electronic band structures of TaAs, TaP, NbAs, and NbP with and without spin-orbit coupling are shown in  Figs.~\ref{fig:bandtaas}, \ref{fig:bandtap}, \ref{fig:bandnbas}, and \ref{fig:bandnbp}, respectively. As expected, the band structures of these four Weyl semimetal compounds are characterized by the $d$ orbitals of Ta or Nb atom, and the $p$ orbitals of As or P atom are similar to one another near the Fermi energy. In the absence of spin-orbit coupling, the valence and conduction bands cross and form closed rings that are bounded on the mirror planes ($k_x=0$ and $k_y =0$ planes), due to the fact that both bands having opposite mirror eigenvalues. For the four Weyl semimetal compounds the nodal rings on the $k_x=0$ mirror planes are shown in Fig.~\ref{fig:loops}. Similarly, the location of nodal rings on the $k_y =0$ mirror planes can be found by applying a $C_4$ rotation. Again, note that away from the mirror planes, which lack the necessary symmetry requirements, accidental band crossings are not found.

In the presence of spin-orbit coupling, the aforementioned nodal rings no longer exist. The valence and conduction bands become fully gapped along the high symmetry lines, as shown in  Figs.~\ref{fig:bandtaas}, \ref{fig:bandtap}, \ref{fig:bandnbas}, and \ref{fig:bandnbp}. The gap opening is smaller when replacing the Ta atom for an Nb atom, which indicates weaker spin-orbit coupling in NbAs and NbP and is consistent with the strong spin-orbit coupling in heavy elements.

The disappearance of nodal rings follows with the appearance of Weyl nodes. Pairs of Weyl nodes around the rings with opposite chirality on the two sides of the mirror planes are found. \cite{huang_weyl_2015,PhysRevX.5.011029} For each ring in  Fig.~\ref{fig:loops}, three pairs of Weyl nodes are generated, and their locations, projected on the $k_x=0$ plane, are indicated by the solid circles. The location of the other pairs of Weyl nodes can be obtained by applying a $C_4$ rotation, which uncovers 24 Weyl nodes for the first Brillouin zone. The detailed $k$-space coordinates and energies of two representative Weyl nodes are given in Table~\ref{table:Weylpoints}. We denote Weyl nodes by $W_1$ for those at $k_z = n$ (multiples of $\frac{2\pi}{c}$), and by $W_2$ otherwise. The energies of $W_1$ are lower than those of $W_2$ for all of the studied compounds. For a Weyl node below (above) the Fermi energy, an electron (hole)- like Fermi surface will enclose a Weyl node, respectively. %The type of Fermi surface can be probed through Hall measurements in transport. However, for the chiral magnetic effect, since the Berry curvature and the Fermi velocity both change signs from a hole-like Fermi surface to an electron-like one, the magneto-current will not depend on the type of the Fermi surface but on its size. \cite{Zhou2013}

Although the nodal rings are destroyed by spin-orbit coupling, new nodal rings are allowed to be created since the closest valence and conduction bands possess opposite mirror eigenvalues, $\pm i$. In other words, bands can cross on the plane with suitable hopping parameters. In NbAs, such a ring is found at $\vec{k}\approx $(0, 0.41, 1.47) with a tiny radius $< 0.005$. The negligible $k$ -space area is in accordance with the energy overlap of only $\leq$0.2 meV for the ring. The energy of the ring is located approximately 5.4 meV above the Fermi energy. Since the energy overlap is tiny and occurs above the Fermi energy, we neglect the significance of this overlap in this study. The fact that spin-orbit coupling can annihilate or keep the nodal lines indicates that these lines depend on the presence of mirror symmetry but are not protected by it. However, parity of the number of nodal lines is conserved because of topology. Since the nodal line is in a $\mathbb{Z}_2$ class ($1+1=0$), \cite{Chiu2014} interaction between two nodal lines can produce zero or two lines.

%\begin{figure}[tbp]
%\includegraphics[width=1.00\columnwidth,clip=true,angle=0]{Fig_Weylform}
%\caption{(Color online). Illustration of how a pair of Weyl nodes can be formed around the $%
%k_x$=0 mirror plane. (a) One part of a ring, line nodes formed by the
%crossing of valence and conduction bands, is shown in the $k_x$=0 mirror
%plane. (b) Several selected bands are plotted along the $k_x$ and $k_y$
%directions, respectively. Each band is doubly degenerated and each $k$ point
%at the line nodes has four-fold degeneracy. (c) After taking spin-orbit
%coupling into consideration, the doubly degenerated bands are split into two
%distinct bands. Two Weyl nodes with opposite values of chirality, which are
%labelled by $+$ and $-$ symbols, respectively, are formed on the opposite
%sides of the $k_x$=0 mirror plane. }
%\label{fig:Weylform}
%\end{figure}

\begin{table}[tbp]
\caption{The coordinates (in units of reciprocal lattice vectors of
conventional unit cell) and energies (in units of eV) of two representative
distinct Weyl nodes denoted as $W_1$ and $W_2$ are given below. In each compound the energy of $W_2$ is higher than that of $W_1$. $\pm$ stands for a mirror pair of Weyl nodes.}
\label{table:Weylpoints}%
\begin{tabular}{ccccc}
\hline\hline
& coordinate of $W_1$ & energy of $W_1$ &  &  \\ \hline
TaAs & ($\pm$0.0072, 0.4827, 1.0000) & -0.0221 &  &  \\
TaP & ($\pm$0.0074, 0.4809, 1.0000) & -0.0531 &  &  \\
NbAs & ($\pm$0.0025, 0.5116, 1.0000) & -0.0322 &  &  \\
NbP & ($\pm$0.0028, 0.5099, 1.0000) & -0.0534 &  &  \\ \hline
& coordinate of $W_2$ & energy of $W_2$ &  &  \\ \hline
TaAs & ($\pm$0.0185, 0.2831, 0.6000) & -0.0089 &  &  \\
TaP & ($\pm$0.0156, 0.2743, 0.5958) & 0.0196 &  &  \\
NbAs & ($\pm$0.0062, 0.2800, 0.5816) & 0.0042 &  &  \\
NbP & ($\pm$0.0049, 0.2703, 0.5750) & 0.0259 &  &  \\ \hline\hline
\end{tabular}%
\end{table}

\subsection{Density of states}

\begin{figure}[tbp]
\includegraphics[width=0.92\columnwidth,clip=true,angle=0]{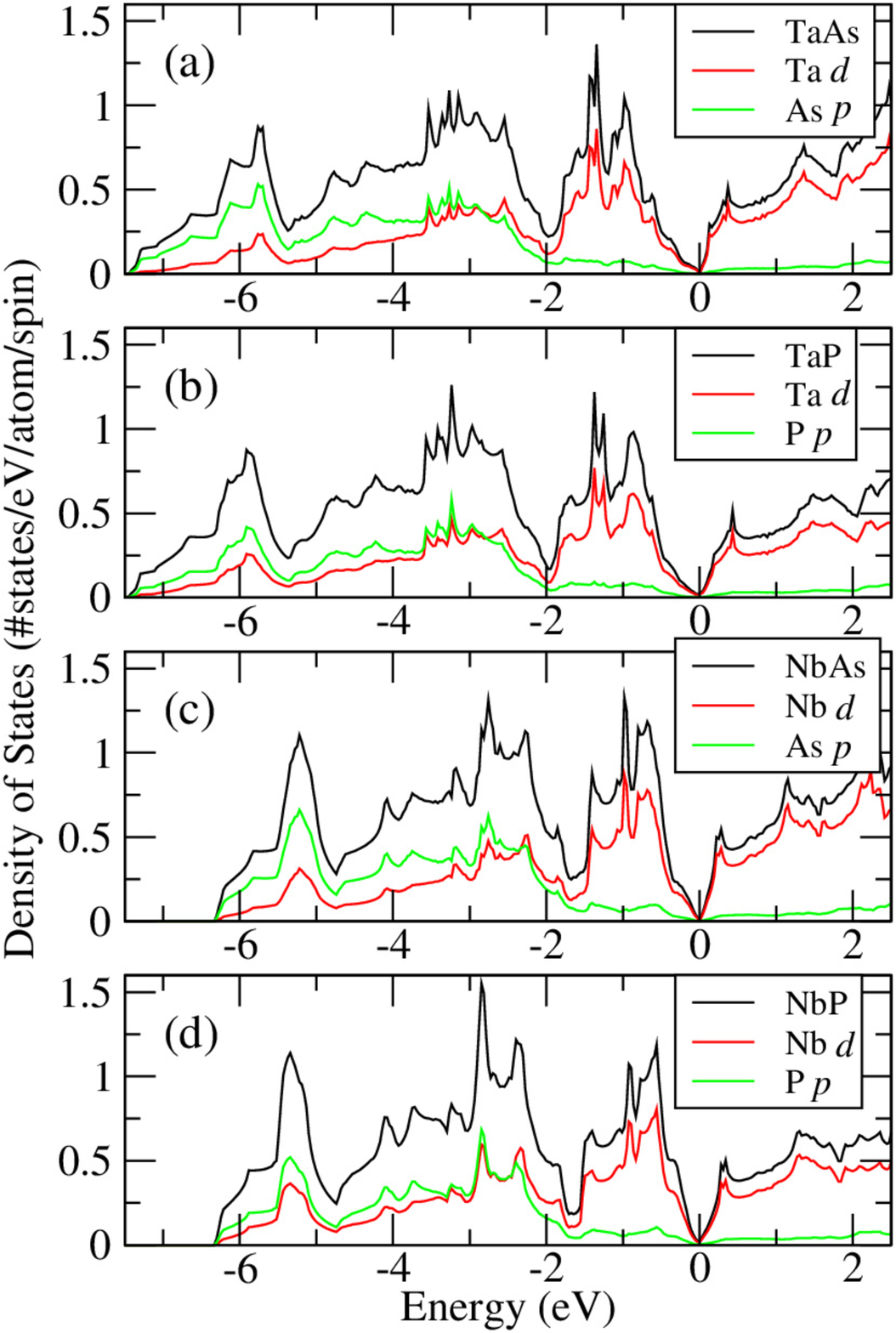}
\caption{(Color online). The orbital contributions of Ta $d$ or Nb $d$ and As $p$ or P $p$
in the density of states with spin-orbit coupling for (a) TaAs, (b) TaP, (c)
NbAs, and (d) NbP are presented. The $d$ and $p$ contributions are colored
red and green, respectively. The total formula contributions are plotted by
black lines. }
\label{fig:dos}
\end{figure}

TaAs, TaP, NbAs, and NbP have similar electron configurations in the outer shells, namely $d^3s^2$ and $p^3$ for Ta, or Nb, and As, or P, respectively. In order to fill the outer-most $p$ orbitals of As and P, As$^{3-}$, P$^{3-}$, Ta$^{3+}$, and Nb$^{3+}$ are expected according to the ionic-bonding picture, which should leave the Ta $d$ or Nb $d$ orbitals having a higher energy
than the full-filled As $p$ or P $p$ orbitals. In Fig.~\ref{fig:dos}, the partial density of states for Ta $d$, Nb $d$, As $p$, and P $p$ are presented. One can directly observe that the $d$ orbitals are strongly hybridized with the $p$ orbitals. Although $p$ character leads to the major contributions in the bottom band, the $p$ orbitals are not fully occupied. In fact, the occupation numbers obtained in the basis of linear combination of atomic orbitals give only approximately half-filled $p$ orbitals and the $p$ orbitals also contribute to the energy above the Fermi energy. Therefore, it is inadequate to neglect the effect of $p$ orbitals near the Fermi energy.

A simple picture to understand the obtained density of states is that of the Ta, or Nb, $d$ orbitals hopping strongly to the As, or P, $p$ orbitals and giving rise to the bonding and anti-bonding bands across the Fermi energy. The Ta, or Nb, $s$ orbital also participates in the hybridization and does not donate itself completely. Having this hybridization scheme, the semimetal feature is formed, which shows the valence and conduction bands separated by a valley shaped density of states at the Fermi energy. The density of states are not zero at the Fermi energy due to the lack of particle-hole symmetry and the two distinct Weyl nodes ($W_1$, $W_2$) possessing different energies for each compound. Furthermore, the valence and conduction bands are found to cut through the Fermi energy and, therefore, forming hole and electron pockets, respectively. Note that the major contribution to the density of states around the Fermi energy is of $d$ character. The bonding and anti-bonding $p$ orbitals are split into much lower and higher energy.

Some additional information can be obtained from Fig.~\ref{fig:dos}. TaAs and TaP share a wider bandwidth relative to the narrower bandwidth of NbAs and NbP, which reflects that 5$d$ electrons are more delocalized than 4$d$ electrons and that the replacement of As $p$ by P $p$ has less of an effect on the overall density of states. However, it is clearly observed that the density of states of P $p$ in TaP is slightly lower than As $p$ in TaAs. In addition, the density of states of P $p$ in NbP is lower than As $p$ in NbAs. This indicates that P $p$ orbitals lose more electrons and are further away from the picture of fully-filled $p$ orbitals. Another observation is that Nb $d$ orbitals in NbP show the largest area below the Fermi energy and, therefore, possess the most $d$ electrons out of the family of four Weyl semimetal compounds.

\section{Charge carriers}

\label{sec:fermisurface}

In this section we investigate the properties of the Fermi surface. We first show the Fermi velocities of the Weyl fermions and the band dispersion between the energies of Weyl nodes and the Fermi energy for $W_1$ and $W_2$. Because of symmetry, we will show results for $W_1$ and $W_2$ at positive $k_x$, listed in Table~\ref{table:Weylpoints}. Then the detailed studies of all the electron- and hole-like Fermi surfaces found in the four Weyl semimetal compounds are discussed.

\subsection{Fermi velocity of Weyl fermion}

\begin{figure}[tbp]
\includegraphics[width=1.00\columnwidth,clip=true,angle=0]{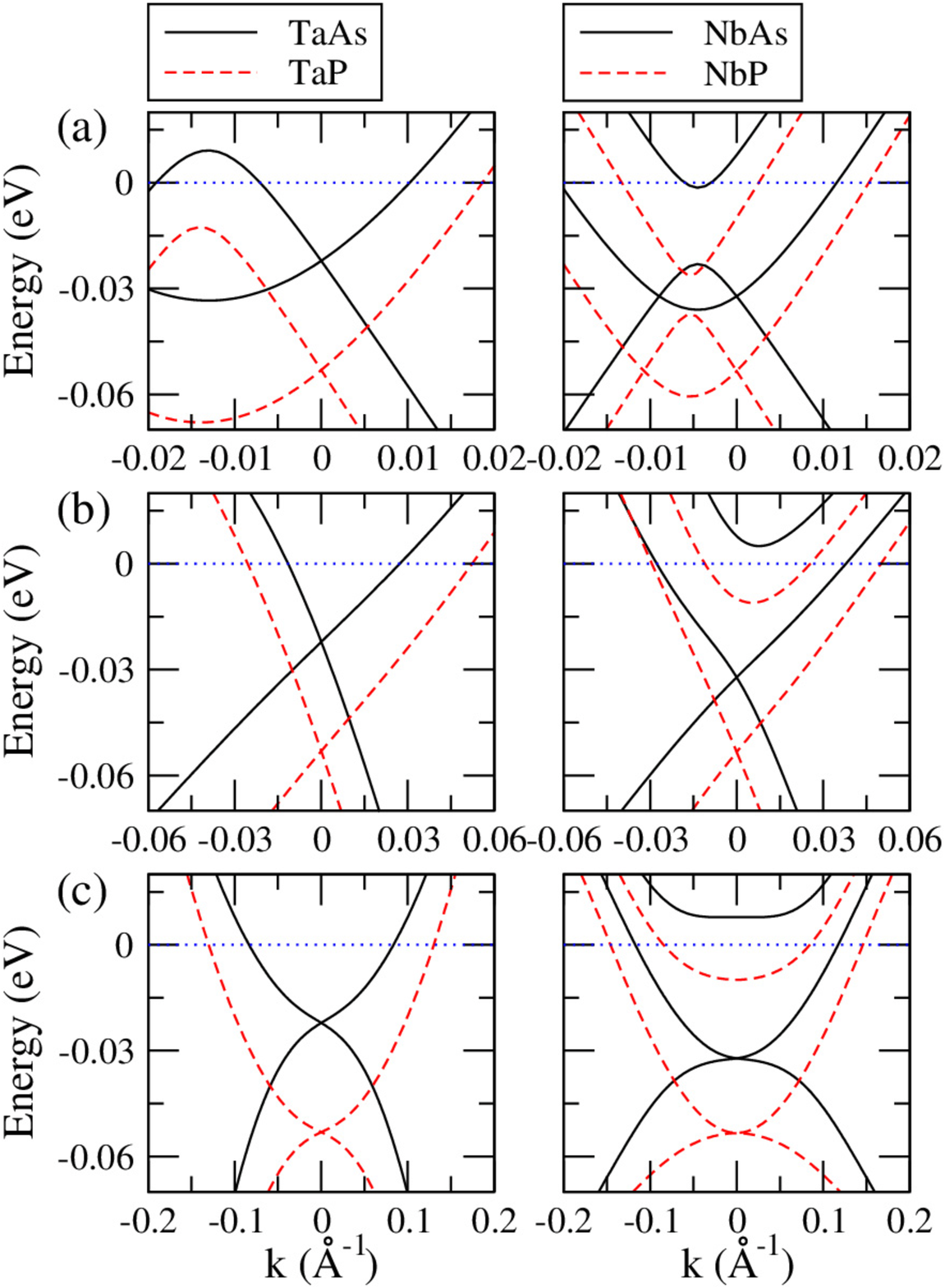}
\caption{(Color online). Band dispersion around $W_1$ along (a) $k_x$, (b) $k_y$, and (c) $%
k_z$ directions of TaAs, TaP, NbAs, and NbP. The $k$-space coordinate of each Weyl node is set to $0$.}
\label{fig:w1dispersion}
\end{figure}

\begin{figure}[tbp]
\includegraphics[width=1.00\columnwidth,clip=true,angle=0]{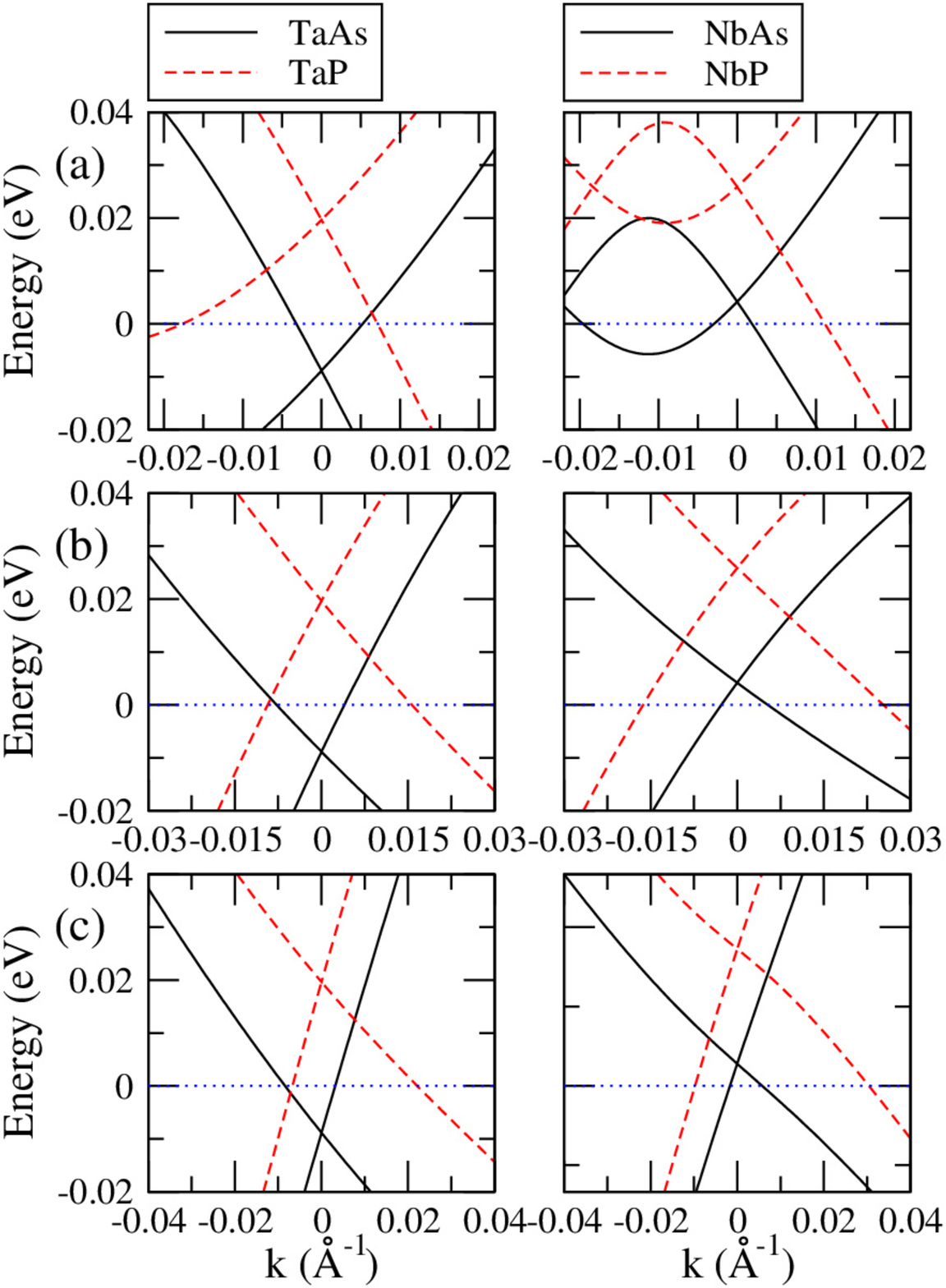}
\caption{(Color online). Band dispersion around $W_2$ along (a) $k_x$, (b) $k_y$, and (c) $%
k_z$ directions of TaAs, TaP, NbAs, and NbP. The $k$-space coordinate of each Weyl node is set to $0$.}
\label{fig:w2dispersion}
\end{figure}

In Weyl semimetals, the energies of Weyl nodes are close to the Fermi energy. Therefore, the physical properties of Weyl fermions are easy to experimentally access by slightly electron or hole doping the systems. The velocity of electron can be calculated by $(1/\hbar) dE/dk$. The details of how the slope of band dispersion varies between the $W_1$ or $W_2$ and the Fermi energy can be found in Fig.~\ref{fig:w1dispersion} and Fig.~\ref{fig:w2dispersion}, respectively. The velocity at $W_1$ and $W_2$  along the $k_x$, $k_y$, and $k_z$ directions are listed in Table~\ref{table:velocity}.  As long as $k$ is close enough to the Weyl nodes, the three-dimensional linear dispersive feature of Weyl cones is expected. It can be found that the velocities in  Table~\ref{table:velocity} are highly anisotropic, and so are the band dispersions shown in Fig.~\ref{fig:w1dispersion} and Fig.~\ref{fig:w2dispersion}. Consequently, the Fermi surface should exhibit such anisotropy. A detailed discussion of the Fermi surfaces will be addressed in the subsequent subsection.

\begin{table}[tbp]
\caption{Mean velocities at Weyl nodes along $x$, $y$, and $z$ directions
in units of $10^{5}\ \text{m} \cdot \text{s}^{-1}$. c and v are for the conduction band and the valence band, respectively.}
\label{table:velocity}%
\begin{tabular}{ccccccc}
\hline\hline
&  &  & TaAs & TaP & NbAs & NbP \\ \hline
$W_1$ & $v_{x}$ & c & 2.5 & 3.1 & 2.5 & 3.7 \\
& $v_{x}$ & v & -5.2 & -5.7 & -4.8 & -5.7 \\
& $v_{y}$ & c & 1.2 & 1.5 & 1.2 & 1.5 \\
& $v_{y}$ & v & -3.2 & -3.6 & -2.0 & -3.0 \\
& $v_{z}$ & c & 0.2 & 0.2 & 0.1 & 0.0(3) \\
& $v_{z}$ & v & -0.2 & -0.2 & -0.1 & -0.0(3) \\ \hline
$W_2$ & $v_{x}$ & c & 2.4 & 2.3 & 2.4 & 2.1 \\
& $v_{x}$ & v & -4.3 & -4.1 & -3.3 & -3.2 \\
& $v_{y}$ & c & 3.5 & 3.0 & 2.3 & 2.1 \\
& $v_{y}$ & v & -1.7 & -2.0 & -1.2 & -1.6 \\
& $v_{z}$ & c & 4.3 & 4.4 & 3.7 & 3.8 \\
& $v_{z}$ & v & -1.6 & -1.5 & -1.1 & -1.0 \\ \hline\hline
\end{tabular}%
\end{table}

\subsection{Fermi surface}

The properties of Fermi surfaces discussed below were calculated using discrete grids in $k$ space. The calculations were performed using a conventional unit cell. Each length of the grid was $0.0015$, $0.0015$, and $0.005$ reciprocal lattice vector along the $k_x$, $k_y$, and $k_z$ direction, respectively. Information regarding the volume and cross-sectional area of Fermi surfaces and the Fermi velocities for each compound are listed in Tabel~\ref{table:concentration}.

\begin{figure}[tbp]
\includegraphics[width=1.00\columnwidth,clip=true,angle=0]{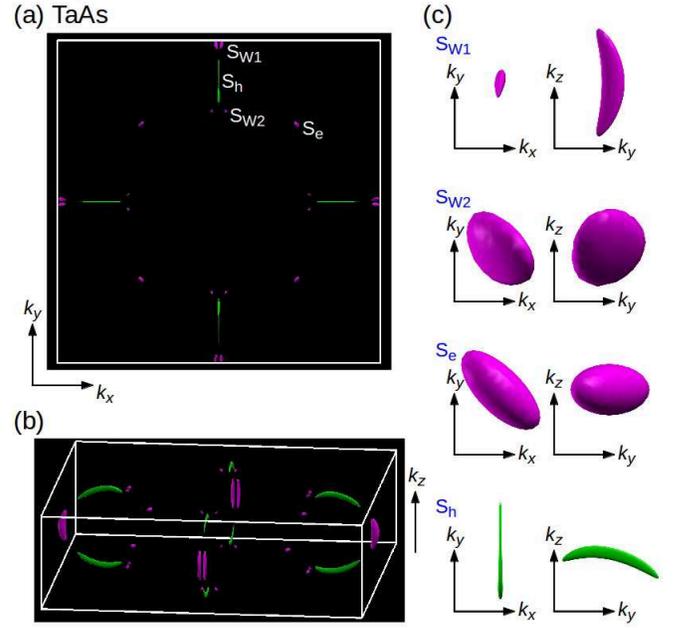}
\caption{(Color online). (a) Top view and (b) bird's eye view of TaAs Fermi surfaces in the
first Brillouin zone of conventional unit cell with (c) the more
detailed plots. Electron-like and hole-like Fermi surfaces are
colored pink and green, respectively. $S_{W1}$ ($S_{W2}$) is the Fermi surface enclosing the $W_1$ ($W_2$) Weyl node(s),
and $S_e (S_h)$ is an electron (hole) pocket without enclosing any Weyl node.}
\label{fig:taasfermisurface}
\end{figure}

\begin{figure}[tbp]
\includegraphics[width=1.00\columnwidth,clip=true,angle=0]{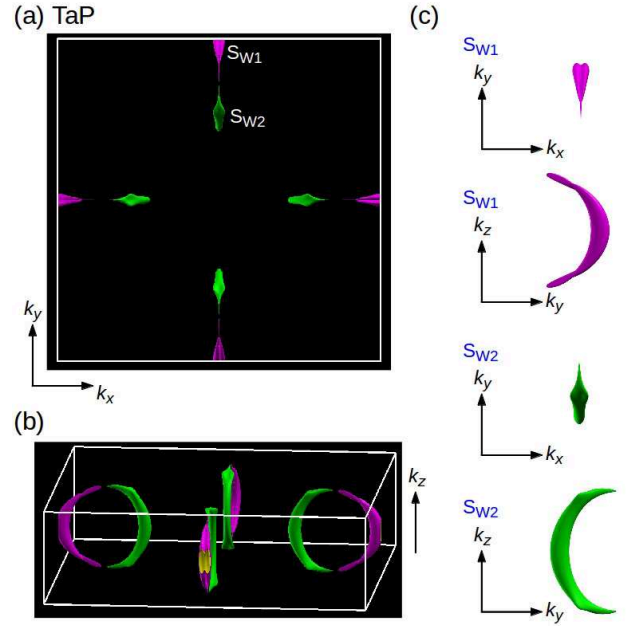}
\caption{(Color online). (a) Top view and (b) bird's eye view of TaP Fermi surfaces in the
first Brillouin zone of conventional unit cell with (c) the more
detailed plots. Electron-like and hole-like Fermi surfaces are
colored pink and green, respectively. $S_{W1}$ ($S_{W2}$) is the Fermi surface enclosing the $W_1$ ($W_2$) Weyl node(s).}
\label{fig:tapfermisurface}
\end{figure}

\begin{figure}[tbp]
\includegraphics[width=1.00\columnwidth,clip=true,angle=0]{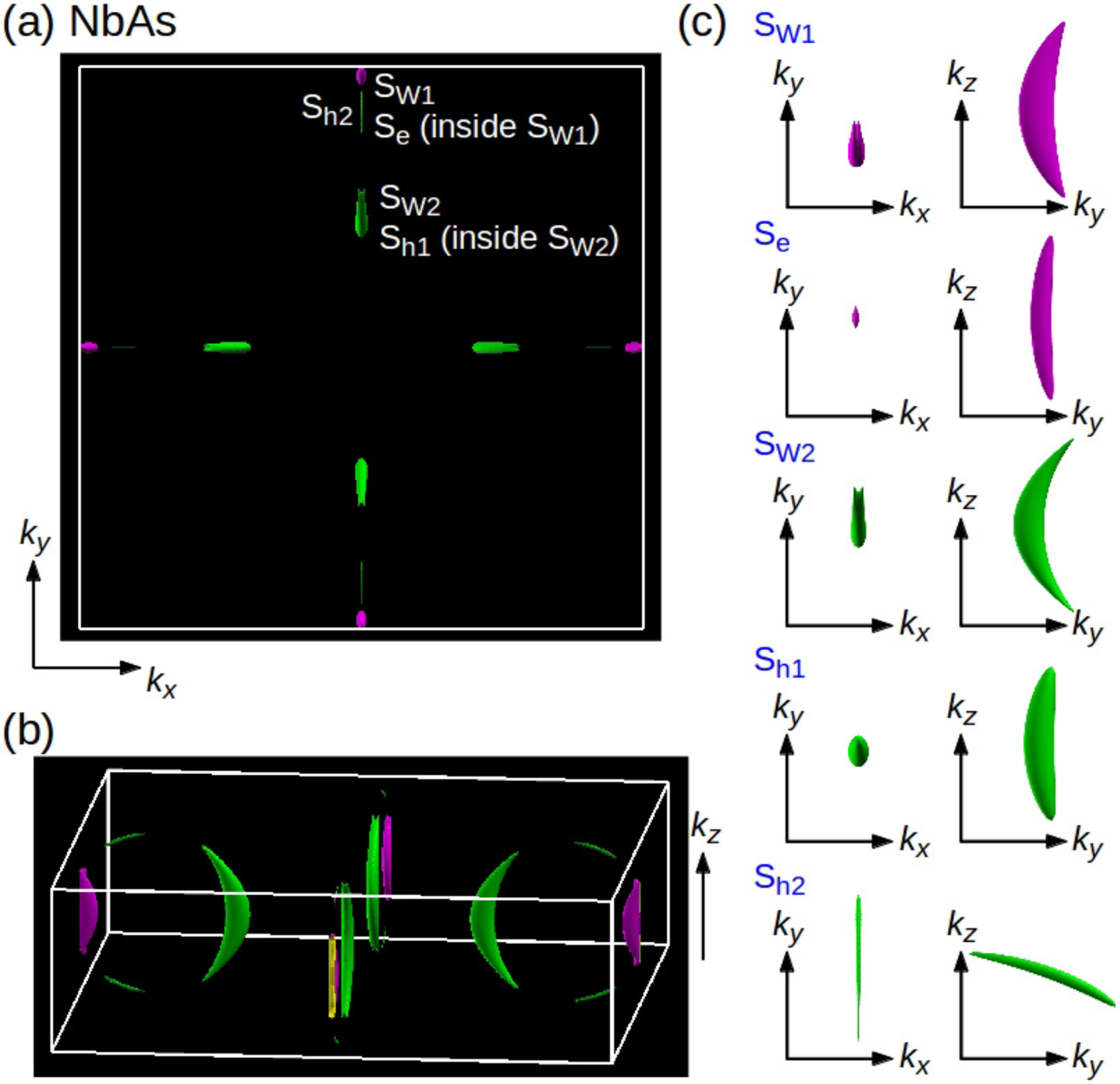}
\caption{(Color online). (a) Top view and (b) bird's eye view of NbAs Fermi surfaces in the
first Brillouin zone of conventional unit cell with (c) the more
detailed plots. Electron-like and hole-like Fermi  surfaces are
colored pink and green, respectively. $S_{W1}$ ($S_{W2}$) is the Fermi surface enclosing and evolved from the $W_1$ ($W_2$) Weyl node(s),
and $S_e (S_h)$ is an electron (hole) pocket not evolved from the Weyl node.}
\label{fig:nbasfermisurface}
\end{figure}

\begin{figure}[tbp]
\includegraphics[width=1.00\columnwidth,clip=true,angle=0]{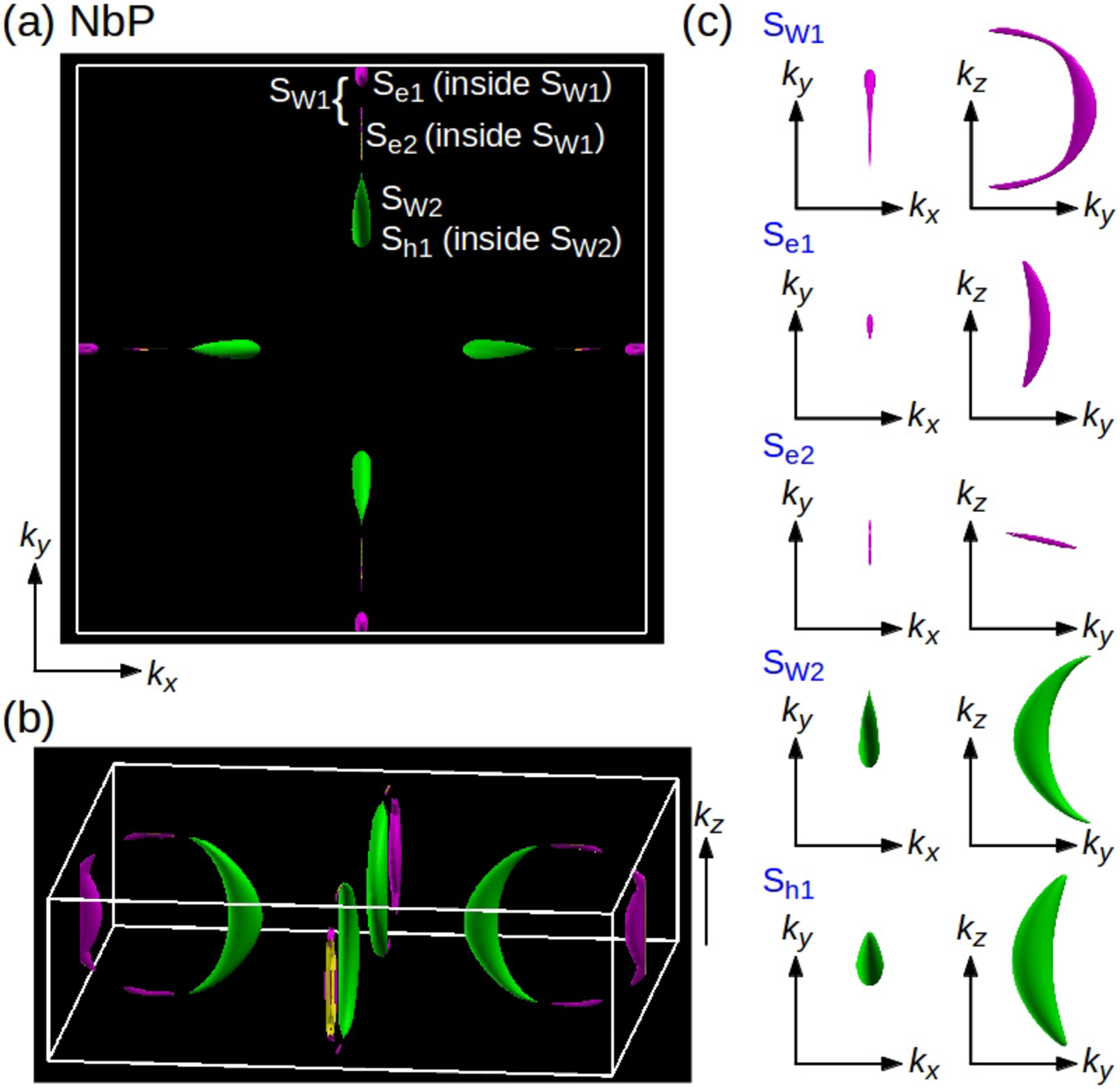}
\caption{(Color online). (a) Top view and (b) bird's eye view of NbP Fermi surfaces in the
first Brillouin zone of conventional unit cell with (c) the more
detailed plots. Electron-like and hole-like Fermi  surfaces are
colored pink and green, respectively. $S_{W1}$ ($S_{W2}$) is the Fermi surface enclosing and evolved from the $W_1$ ($W_2$) Weyl node(s),
and $S_e (S_h)$ is an electron (hole) pocket not evolved from the Weyl node.}
\label{fig:nbpfermisurface}
\end{figure}

\subsubsection{TaAs}

The Fermi surfaces of TaAs are formed by both electron and hole pockets, which are shown in Fig.~\ref{fig:taasfermisurface}. Ac- cording to Table~\ref{table:Weylpoints}, both $W_1$ and $W_2$ are below the Fermi energy, with the latter being closer to it. As a result, an electron pocket enclosing $W_1$ or $W_2$ Weyl node is expected. The Fermi surface of $W_1$ is much bigger than that of $W_2$ and shows high anisotropy along $k_z$. Some distance away from the end of the Fermi surface of $W_1$, a hole pocket centered on the mirror plane appears and extends to the neighborhood of the Fermi surface of $W_2$. Since each of the $S_{W1}$ and $S_{W2}$ Fermi surfaces encloses one Weyl node, the Fermi surface is endowed with a $\pm1$ Chern number.\cite{2015arXiv150507727G} Furthermore, the hole pocket does not surround any Weyl nodes, which is an outcome of the modulation of bands, and, consequently, carries a zero Chern number. The shape of the Fermi surfaces is reminiscent of the nodal rings in the absence of spin-orbit coupling. We note that different from the other compounds discussed below, the Fermi surfaces on two sides of a mirror plane are isolated instead of getting immersed, which is due to the distances between Weyl nodes and their energies relative to the Fermi energy. Furthermore, an electron pocket around  (0.2425, 0.2425, 0) is found, which does not appear in other compounds. The details of the shapes and carrier concentrations can be found in Fig.~\ref{fig:taasfermisurface} and Table~\ref{table:concentration}.  The maximal areas of the cross sections for the Fermi surfaces, which can be specifically measured by cyclotron experiments, are also listed in Table~\ref{table:concentration}.

\subsubsection{TaP}

The Fermi surfaces of TaP are presented in Fig.~\ref{fig:tapfermisurface}. Contrary to TaAs, in which both energies of $W_1$ and $W_2$ are below the Fermi energy, the energy of $W_1$ and $W_2$ in TaP are below and above the Fermi energy, respectively shown in Table~\ref{table:Weylpoints}. The $W_2$ in TaP is enclosed by a hole pocket Fermi surface. Due to higher energies of Weyl nodes away from the Fermi energy, the volume of Fermi surfaces are much larger (about $10\sim20$
 times for $S_{W1}$) than those in TaAs. Furthermore, a Fermi surface encloses more than one Weyl node. An electron-like Fermi surface encompasses a mirror pair of $W_1$ points, while a hole-like Fermi surface covers two mirror pairs of $W_2$ points. Compared to that in TaAs, both electron and hole pockets take the shape of a crescents that are distributed more along the $k_z$ direction. Except these electron and hole pockets there is no other piece of Fermi surfaces. The carrier concentrations and maximal areas of cross sections are listed in Table~\ref{table:concentration}.

\subsubsection{NbAs}

The Fermi surfaces of NbAs are presented in Fig.~\ref {fig:nbasfermisurface}. In NbAs, the energy of $W_1$ and $W_2$ are below and above the Fermi energy, respectively. Three types of hole-like Fermi surface are found in NbAs. The first kind encloses four $W_2$ points and passes through the $k_z=0$ plane, like those found in TaP. The second kind is similar to those crescents elongated along the $k_x$ or $k_y$, direction in TaAs. The third kind cannot be seen in Fig.~\ref{fig:nbasfermisurface} because the Fermi energy cuts through two spin split valence bands at $\vec{k}\approx$(0, 0.2152, 0). Consequently, the inner surface is hidden inside the outer surface. In comparison with the $W_1$ points in TaAs and TaP, the two neighboring $W_1$ points are closer to each other. As a result, only one big crescent-shaped electron pocket containing two $W_1$ points is observed. Similar to the scenario of the two concentric-like hole pockets, there is also another electron pocket hidden inside the bigger Fermi surface near the $W_1$. The carrier concentrations and maximal areas of cross sections for each Fermi surface are listed in Table~\ref{table:concentration}. 

\begin{table}[tbp]
\caption{Electron ($e$) and hole ($h$) carrier concentrations ($n$), maximal
areas ($A$) of cross sections on the $yz$, $xz$, and $xy$ planes, and
root-mean-square velocities ($\bar{v}$) along the $k_x$, $k_y$, and $k_z$ directions
of Fermi surfaces ($cf.$ Fig.~\protect\ref{fig:taasfermisurface}). $S_{W1} (S_{W2})$ is the Fermi surface enclosing and evolved from the $W_1(W_2)$ Weyl node(s), and $S_e (S_h)$ is an electron (hole) pocket Fermi surface not evolved from any Weyl node. The units
of carrier concentration, area, and velocity are $10^{17}\ \textrm{cm}^{-3}$, $%
10^{-3}\ \text{\AA }^{-2}$, and $10^{5}\ \text{m} \cdot \text{s}^{-1}$, respectively.}
\label{table:concentration}%
\begin{tabular}{ccccccccc}
\hline\hline
TaAs &          & $n$ & $A_{yz}$ & $A_{xz}$ & $A_{xy}$ & $\bar{v}_x$ & $\bar{v}_y$ & $\bar{v}_z$ \\
\hline
& $S_{W1} (e)$ & $3.336$ & $6.060$ & $2.557$ & $0.639$ & $2.920$ & $1.532$ & $0.453$ \\
& $S_{W2} (e)$ & $0.134$ & $0.244$ & $0.237$ & $0.226$ & $2.055$ & $1.579$ & $1.763$ \\
& $S_{e}$ & $0.364$ & $0.406$ & $0.406$ & $0.624$ & $0.365$ & $0.365$ & $0.412$ \\
& $S_{h}$ & $2.201$ & $7.332$ & $0.310$ & $1.639$ & $4.051$ & $0.670$ & $1.372$ \\
\hline
TaP &          & $n$ & $A_{yz}$ & $A_{xz}$ & $A_{xy}$ & $\bar{v}_x$ & $\bar{v}_y$ & $\bar{v}_z$ \\
\hline
& $S_{W1} (e)$ & $48.98$ & $23.67$ & $15.87$ & $4.687$ & $3.270$ & $1.594$ & $0.767$ \\
& $S_{W2} (h)$ & $53.17$ & $36.96$ & $13.94$ & $3.404$ & $3.736$ & $2.542$ & $1.879$ \\
\hline
NbAs &          & $n$ & $A_{yz}$ & $A_{xz}$ & $A_{xy}$ & $\bar{v}_x$ & $\bar{v}_y$ & $\bar{v}_z$ \\
\hline
& $S_{W1} (e)$ & $14.60$ & $15.97$ & $6.882$ & $1.752$ & $3.750$ & $1.370$ & $0.542$ \\
& $S_{W2} (h)$ & $29.76$ & $24.57$ & $10.29$ & $2.646$ & $4.647$ & $2.647$ & $1.765$ \\
& $S_{e}$ & $0.196$ & $1.704$ & $0.375$ & $0.060$ & $2.331$ & $1.131$ & $0.194$ \\
& $S_{h1}$ & $1.925$ & $3.533$ & $2.299$ & $0.470$ & $2.815$ & $1.922$ & $0.600$ \\
& $S_{h2}$ & $0.155$ & $1.410$ & $0.037$ & $0.104$ & $3.692$ & $0.695$ & $1.887$ \\
\hline
NbP &          & $n$ & $A_{yz}$ & $A_{xz}$ & $A_{xy}$ & $\bar{v}_x$ & $\bar{v}_y$ & $\bar{v}_z$ \\ \hline
& $S_{W1} (e)$ & $32.64$ & $31.47$ & $10.78$ & $2.805$ & $4.047$ & $1.653$ & $0.932$ \\
& $S_{W2} (h)$ & $82.16$ & $43.94$ & $20.15$ & $5.913$ & $4.964$ & $3.147$ & $2.551$ \\
& $S_{e1}$ & $4.323$ & $11.12$ & $2.928$ & $0.599$ & $4.773$ & $1.845$ & $0.501$ \\
& $S_{e2}$ & $0.025$ & $0.219$ & $0.016$ & $0.056$ & $3.828$ & $0.466$ & $1.896$ \\
& $S_{h1}$ & $33.95$ & $22.59$ & $13.03$ & $3.772$ & $4.971$ & $3.294$ & $1.929$ \\
\hline\hline
\end{tabular}%
\end{table}

\subsubsection{NbP}

The Fermi surfaces of NbP are presented in Fig.~\ref{fig:nbpfermisurface}. Similar to TaP and NbAs, the energy of $W_1$ and $W_2$ in NbAs are below and above the Fermi energy, respectively. The two concentric-like hole Fermi surfaces split by spin-orbit coupling are also present in NbP as intersecting the the $k_z$=0 plane at $\vec{k}\approx$(0, 0.2042, 0), giving rise to four crescent Fermi surfaces in the first Brillouin zone with each surface containing one more surface inside. Each outer surface encloses one pair of $W_2$ points. In NbP, one long crescent electron-like Fermi surface resembling the shape of half a ring can be found. The surface becomes thinner while deviating from the $W_1$ and the detailed shape is easier to be observed in Fig.~\ref{fig:nbpfermisurface}(c). There are two additional electron-like Fermi surface inside this long crescent surface, $S_e1$ and $S_e2$. Referring to the band dispersion in Fig.~\ref{fig:w1dispersion}, we have to carefully identify the concentric electron pockets. Although $S_e1$ encloses one W1 point, the associated band does not evolve from the band of Weyl nodes and, therefore, is a trivial Fermi surface. The size of the surface $S_e2$ is small and touches the limit for the chosen grid size. The carrier concentration  and maximal areas of cross sections for each Fermi surface is listed in Table~\ref{table:concentration}.

\section{Discussion}

\label{sec:discussion}

Several interesting properties of the charge carriers can be found even without the consideration of additional conditions, like applying magnetic or strain fields in first-principles calculations. We first note that the energy difference between $W_1$ and $W_2$ is $\sim 10 - 100$ meV, which is realistic to access by doping in real crystals. The Fermi energy for TaAs was experimentally found \cite{2015arXiv150302630Z} to be located between $W_1$ and $W_2$ ($E_{W1}$ and $E_{W2}$, respectively), which is approximately 11.5 meV above $E_{W1}$ and, therefore, different from our calculations shown in Table~\ref{table:Weylpoints}. %Other distinct types of charge carriers can be modulated as a case can show both hole pockets enclosing $W_1$ and $W_2$.

The energy difference between two Weyl nodes is directly related to the minimal area of the Fermi surface as a function of binding energy, with a larger separation, $|{E_{W2} - E_{W1}}|$, corresponding to a larger minimal Fermi surface. Referring to Table~\ref{table:Weylpoints}, the energy differences of $E_{W2}$ and $E_{W1}$  are 0.0132, 0.0727, 0.0364, and 0.0793 eV
for TaAs, TaP, NbAs, and NbP, respectively.  Recall, $E_{W2} > E_{W1}$. Note that while the separation in momentum space of the Weyl nodes is determined by the strength of the spin-orbit coupling, the separation in energy of the Weyl nodes can be viewed as arising from the dispersion of the spinless line node crossing, observed before spin-orbit coupling is introduced. Amongst the four Weyl semimetal compounds, the relatively small value for TaAs explains the relatively small carrier concentration, e.g. 0.134$\times10^{17}\ \textrm{cm}^{-3}$ for $S_{W2}$. Similarly, the largest hole pocket can be found in NbP, e.g. 82.16$\times10^{17}\ \textrm{cm}^{-3}$ for $S_{W2}$. The rich and tunable behavior of charge carriers in these four semimetals strongly suggests that these compounds are a good playground for engineering the electronic properties through doping, strain, electric fields, magnetic fields and/or other external perturbations.

Furthermore, we also point out that an abrupt change in the Fermi surface volume may occur by tuning, for example, the Fermi energy. Since each pair of Weyl nodes are located on two sides of the mirror plane, after turning on the spin-orbit coupling, they are, in general, close to each other.  Therefore, there exists a critical Fermi energy that can make the two Fermi surfaces,  each enclosing the Weyl node(s) on one side of the plane, touch at one $k$ point. By changing the critical Fermi energy, either one big surface enclosing both Weyl nodes or two small surfaces with each enclosing a single Weyl node can be formed. The Fermi surfaces of the $W_1$ or $W_2$ pairs in TaAs is an example for the latter case while those in TaP, NbAs, and NbP are examples of the former case.

\end{document}